\documentclass[epj]{svjour}

\usepackage{graphicx}
\usepackage{fancyhdr}
\usepackage{amssymb}
\def\makeheadbox{{
 }}
\def\mytitle{My title} 
\def\myauthors{My name}  
\def\mytype{My type of session}
\def\mysession{My session}

\def\nt{\tilde{\chi}^0}
\def\stau{\tilde{\tau}}
\def\tb{\tan\beta}
\def\surn{SU(5)_{\rm RN}}
\def\relic{\Omega_{\rm DM }} 
\def\mgut{M_{\rm GUT}}

\def\mytitle{SUSY-GUTs and neutralino DM} 
\def\myauthors{Lorenzo Calibbi}    
\def\mytype{Contributed Talk}    
\def\mysession{Cosmology and Astrophysics}

\begin{document}
\title{Neutralino Dark Matter in SUSY-SU(5) with RH neutrinos}
\author{Lorenzo Calibbi
\thanks{\emph{Email:} lorenzo.calibbi@uv.es}%
}                     
%
%
\institute{SISSA/ISAS, I-34013 Trieste, Italy
\and
Departament de F\'{\i}sica Te\`orica, Universitat de 
Val\`encia-CSIC, E-46100, Burjassot, Spain.
}
%
\date{}
\abstract{
In the context of low-energy Supersymmetry (SUSY), a model of Grand Unification (GUT) with 
right-handed (RH) neutrinos, based on the group $SU(5)$, is discussed and its implications for
neutralino dark matter (DM) are studied and compared with the constrained MSSM. 
RG effects in this model modify the sparticle spectrum such that the WMAP limit on the
DM relic density cannot
be satisfied for small values of $\tb$ ($\tb \lesssim 35$) and the region of the parameter space
allowed by efficient neutralino-stau coannihilation presents a peculiar phenomenology 
and an upper bound on the neutralino mass for most of the parameters choices. 
\PACS{
      {12.60.Jv}{Supersymmetric models}   \and
      {95.35.+d}{Dark matter}
     } 
} 
\maketitle
\section{Introduction: neutralino DM in CMSSM}
\label{intro}

In the minimal supersymmetric extension of the Standard Model (MSSM), imposition of R-parity  
makes the lightest supersymmetric particle (LSP) stable. 
Therefore, a neutral and uncolored LSP would be a good candidate to explain dark matter 
in terms of WIMP (Weakly Interacting Massive Particle). 
This is the case of several SUSY models, where the LSP is the lightest neutralino ($\nt_1$).   

Thanks to the recent WMAP results \cite{wmap}, the DM relic density is very well known
and this translates into very strict constraints on the parameter space of the model.
For instance in the Constrained MSSM (CMSSM), where the so-called universality condition 
is imposed at the GUT scale, $\nt_1$ is usually bino-like, thus very weakly
interacting. This makes the annihilation cross-section of the LSP very small and the DM relic 
density $\relic$ easily exceed the WMAP limit (here at 3$\sigma$):
\begin{equation}
0.087\lesssim\relic h^2\lesssim 0.138
\label{eq:WMAP}
\end{equation}
There are only three regions
of the parameter space\footnote{The parameters of the model are the universal soft SUSY breaking
parameters (the common scalar mass $m_0$,
gaugino mass $M_{1/2}$ and trilinear coupling $A_0$) and, after the requirement
of electro-weak symmetry breaking, $\tb$ and the sign of the Higgs bilinear coupling $\mu$.} 
which provide the correct relic density for neutralino DM and are not
excluded by the LEP limits on the mass of the SUSY particles: (i) the $\stau$ coannihilation region 
\cite{staucoann}, (ii) the A-pole funnel region \cite{polefunnel} and (iii) the Focus point \cite{focuspoint}.   
In such regions the (co)annihilation cross-section of $\nt_1$ is enhanced and the DM relic density
doesn't exceed the WMAP bound. 
The regions listed above respectively mean: 
(i) effective $\stau_1$-$\nt_1$ coannihilation when $m_{\stau_1}\simeq m_{\nt_1}$;
(ii) resonant enhancement of the LSP annihilation via the s-channel mediation of the CP-odd Higgs $A$ 
for $m_A \simeq 2 m_{\nt_1}$ (which requires large values of $\tan\beta$);
(iii) enhancement of the Higgsino component of $\nt_1$ (and so of the annihilation cross-section)
for small values of $\mu$.   
As we can see, all these WMAP-allowed regions require very special relations among the parameters
and moreover two of them, namely $\stau$ coannihilation and Focus Point, 
lie close to regions of the parameter space excluded by theoretical 
requirements: respectively, neutral LSP, i.e. $m_{\nt_1} < m_{\stau_1}$, 
and viable radiative electro-weak symmetry breaking (REWSB), i.e. $\mu^2 > 0$.
This is an important point, as we'll see in the following discussion.

In the present talk, we will consider the possibility that extensions of the CMSSM, 
namely an evolution of the parameters above the GUT scale $\mgut$ and/or the presence of
heavy sterile RH neutrinos, destabilize the critical relations
among parameters needed to satisfy the WMAP constraint and thus modify the phenomenology
of neutralino DM.

\section{SUSY-$\surn$: framework and parameter space}

In this section, we will present a simple SUSY-GUT framework, based on $SU(5)$ with
the addition of RH neutrino fields. The MSSM superfields are embedded in the $10$
and $\bar{5}$ representation of $SU(5)$, while the Higgs doublets sit in two
five-dimensional representation that we will call $5_u$ and $\bar{5}_d$. RH neutrinos
are singlets of $SU(5)$.
The resulting superpotential, omitting generation indices, reads:

\begin{eqnarray}
\label{superp}
W_{\surn}& =  & Y^u 10 ~10 ~5_u + Y^d 10 ~\bar{5}~ \bar{5}_d
+ Y^\nu \bar{5}~ 1~ 5_u + \nonumber\\
& &  M_R 1~ 1 + \mu~ 5_u~ \bar{5}_d
\end{eqnarray}
where $Y^u$, $Y^d$ and $Y^\nu$ are the Yukawa couplings, $M_R$ a Majorana mass matrix for RH neutrinos\footnote{
In the following, we will consider just one RH neutrino, while for a consistent generation of neutrino masses
at least two are needed.} and $\mu$ the bilinear Higgs coupling. Our hypothesis is that 
the neutrino Yukawa coupling $Y^\nu$ is as large as the Yukawa coupling of the quark top. Such hypothesis
is natural when the model is incorporated in a $SO(10)$ SUSY-GUT \cite{masiero-vempati}. 

\begin{figure}
\begin{center}
\includegraphics[width=0.45\textwidth]{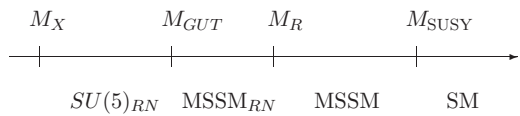}
\caption{Energy scales involved in the model.}
\label{scales}   
\end{center}
\end{figure}

If the the universality condition is imposed at a scale $M_X\approx 5\cdot10^{17}~{\rm GeV}$ above 
$\mgut\approx 2\cdot 10^{16}~{\rm GeV}$ (see Fig.~\ref{scales}), and the relevant superpotential above
$\mgut$ is Eq.~(\ref{superp}), we expect a drastic change in the renormalization group (RG) evolution
of the SUSY parameters from high energy down to the low scale. In particular, $\surn$ presents two
new effects with respect to CMSSM:
(i) a running of the soft scalar masses above $\mgut$, driven by the $\surn$ RGEs;
(ii) the contribution of the neutrino Yukawa interactions to the RG evolution of the soft masses,
in particular for the LH sleptons.

As we mentioned above, in the CMSSM a significant part of the parameter space is ruled out, in one
case because the LSP turns out to be $\stau_1$, in another case because REWSB doesn't take place 
correctly. An interesting consequence of the modification of the RG running in $\surn$ is 
getting rid to both these excluded regions, opening up the parameter space \cite{cfmv,cmv}. 
Since in the CMSSM two of the three DM branches listed in the previous section have a connection with 
these excluded regions, we can expect that $\surn$ significantly changes the $\nt_1$ DM phenomenology.
 
The disappearance of `$\stau$-LSP' and `no-REWSB' regions can be easily understood having a look
at the effects of the RG running in $\surn$ \cite{cfmv,cmv}.  
The right-handed slepton $\stau_R$ sits in the 
$10$ of $SU(5)$ and thus, in the running from the universality scale $M_X$ down to $\mgut$, 
its mass receives contributions from the full unified gaugino 
sector. At the leading log level, this contribution is given at the GUT scale as:
\begin{equation}
m_{\stau_R}^2(\mgut)~ \simeq~{144 \over 20 \pi}\alpha_5 M_{1/2}^2  
\ln({M_X \over \mgut}) ~\simeq ~0.25~ M_{1/2}^2 
\label{pregut}
\end{equation}
where the limit $m_0 \to 0$ is taken and $\alpha_5= g_5^2/4\pi$ 
represents the unified gauge coupling at $\mgut$. This large positive
contribution to the $\tilde{\tau}_R$ mass makes the stau heavier
than $\nt_1$ for most of the parameter space.

Coming to the REWSB issue, let us consider the standard (tree-level)
relation coming from the minimization of the Higgs potential \cite{rewsb}:
\begin{equation}
\mu^2 = \frac{-m_{H_u}^2 \tan^2\beta + m_{H_d}^2}{\tan^2\beta -1} 
-\frac{1}{2}M_Z^2 
\label{electroweak}
\end{equation}
where $m_{H_u}$ and $m_{H_d}$ are the soft masses of the up-type and down-type Higgs doublets.
Eq.~(\ref{electroweak}) is usually used to fix $\mu^2$ letting $\tan\beta$ as a free parameter.
If $\mu^2$ turns out to be negative, the corresponding point of the parameter space cannot
give a viable vacuum. It's clear that a sufficient condition for REWSB to take place is that $m_{H_u}^2$  
(which is equal to $m_0^2$ at $M_X$) gets a negative value at low energy as an effect of
the RG evolution. In CMSSM this is achieved thanks to the large negative corrections
driven by the $\mathcal{O}(1)$ top Yukawa coupling, $y_t$. In presence of RH neutrinos,
the RGE for $m_{H_u}^2$ reads (neglecting gaugino contributions) \cite{petcov-profumo}:
\begin{eqnarray}
 (4\pi)^2 {\partial  m_{H_u}^2 \over \partial \ln (\tilde \mu/M_X)} & \simeq &
6 y_t^2 (m_{H_u}^2 + m_{\tilde{U}_3}^2 + m_{\tilde{Q}_3}^2 + A_t^2) +\nonumber \\
&&  2 y_{\nu}^2 (m_{H_u}^2 + m_{\tilde{N}}^2 + m_{\tilde{L}_3}^2 + A_{\nu}^2)
\label{ynurunning}
\end{eqnarray}
where $m_{\tilde{Q}_3}^2$ and $m_{\tilde{U}_3}^2$ are the 
$\tilde{t}_L$ and $\tilde{t}_R$ soft masses, 
while $m_{\tilde{L}_3}^2$, $m_{\tilde{N}}^2$ are the same for 
the LH and RH sneutrinos respectively. As we can see, our assumption
of a top-like neutrino Yukawa $y_\nu$ adds a further large negative contribution
to the renormalization of $m_{H_u}^2$. Moreover in $SU(5)$  
the squark masses  $m_{\tilde{U}_3}^2$ and $m_{\tilde{Q}_3}^2$ are increased 
by the unified gauge sector
running, such as $m_{\stau_R}^2$ in Eq.~(\ref{pregut}). These two effects of $\surn$
contribute to make easier to achieve $m_{H_u}^2<0$ and thus a correct REWSB, getting
rid to the `no-REWSB' region usually present in the CMSSM parameter space.    

In the next section, we will discuss the phenomenology of neutralino DM in $\surn$,
presenting the results published in Ref.~\cite{cmv}. 

\section{Neutralino DM in $\surn$}

\begin{figure}[t]
\begin{center}
\includegraphics[width=0.35\textwidth]{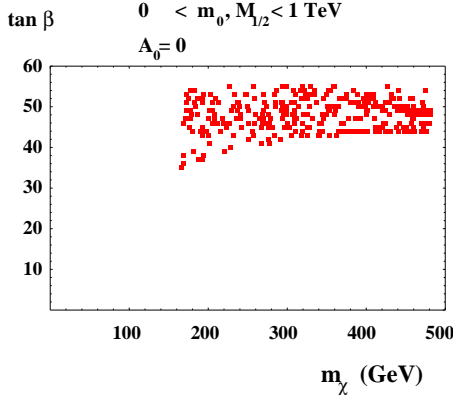}
\end{center}
\caption{Points allowed by experimental and theoretical constrains
after a scan on  $0<m_0,\,M_{1/2}<1~\mbox{TeV}$. 
\label{tb-mchi} }
\end{figure}
 
The modifications of the $\surn$ parameter space with respect to CMSSM are sufficient
to offset the conditions which give viable dark matter in CMSSM at low $\tan\beta$. 
In Fig.~\ref{tb-mchi}, we plot
all the points which satisfy the available direct/indirect constraints 
and give viable relic density as a function of $\tan\beta$ and LSP 
mass. From the figure we see that viable DM is only possible for values
of:
\begin{equation}
\tan\beta~\gtrsim~34 ~;~~~~ m_{\nt_1}~\gtrsim 160~ {\rm GeV}
\label{lowerbounds}
\end{equation}
These are quite strong \textit{lower bounds} on the neutralino mass 
and $\tan\beta$ and will be useful in distinguishing this model compared
to the standard CMSSM parameter space. A remarkable point is that 
such lower bounds remain valid even choosing non-vanishing values of $A_0$, 
up to $\left|A_0\right|\simeq 3\, m_0$ (larger values are excluded by the arising of
tachyonic $\stau$ in most of the parameter space).

\begin{figure}[t]
\begin{center}
\includegraphics[width=0.36\textwidth]{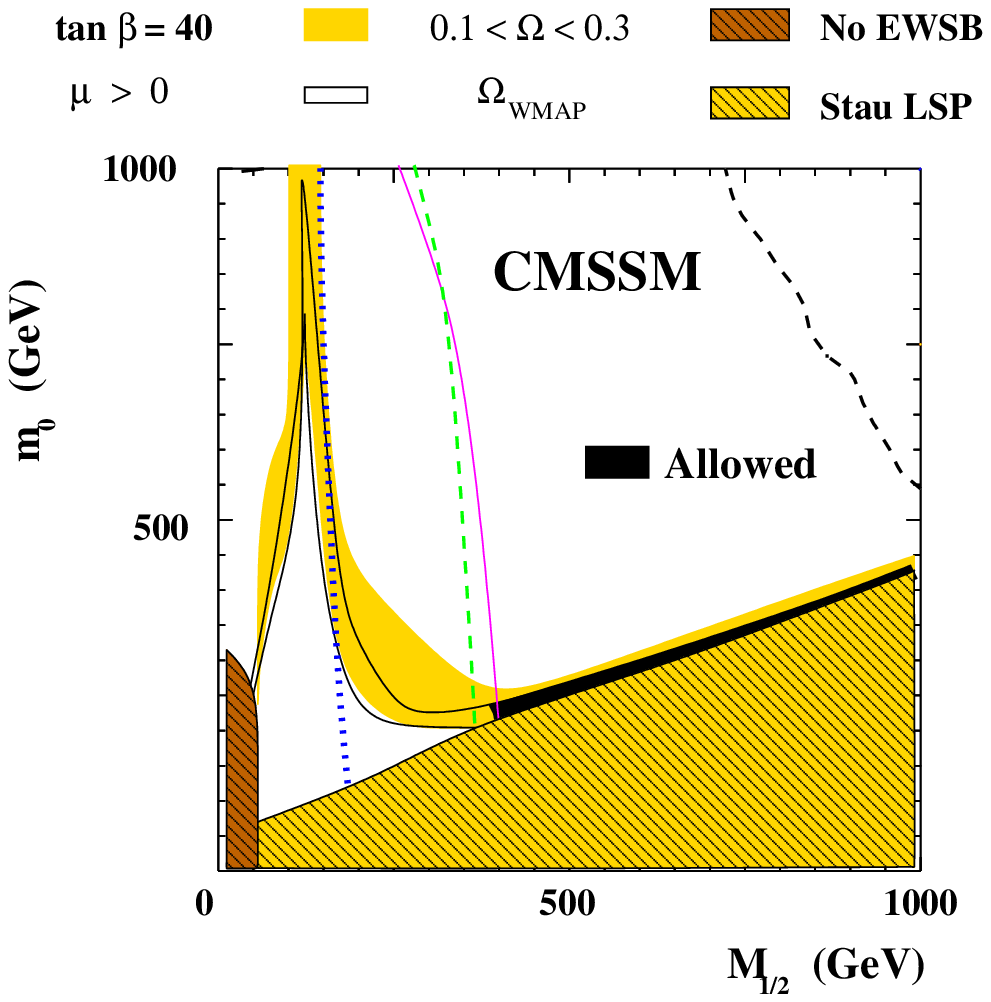}
\includegraphics[width=0.36\textwidth]{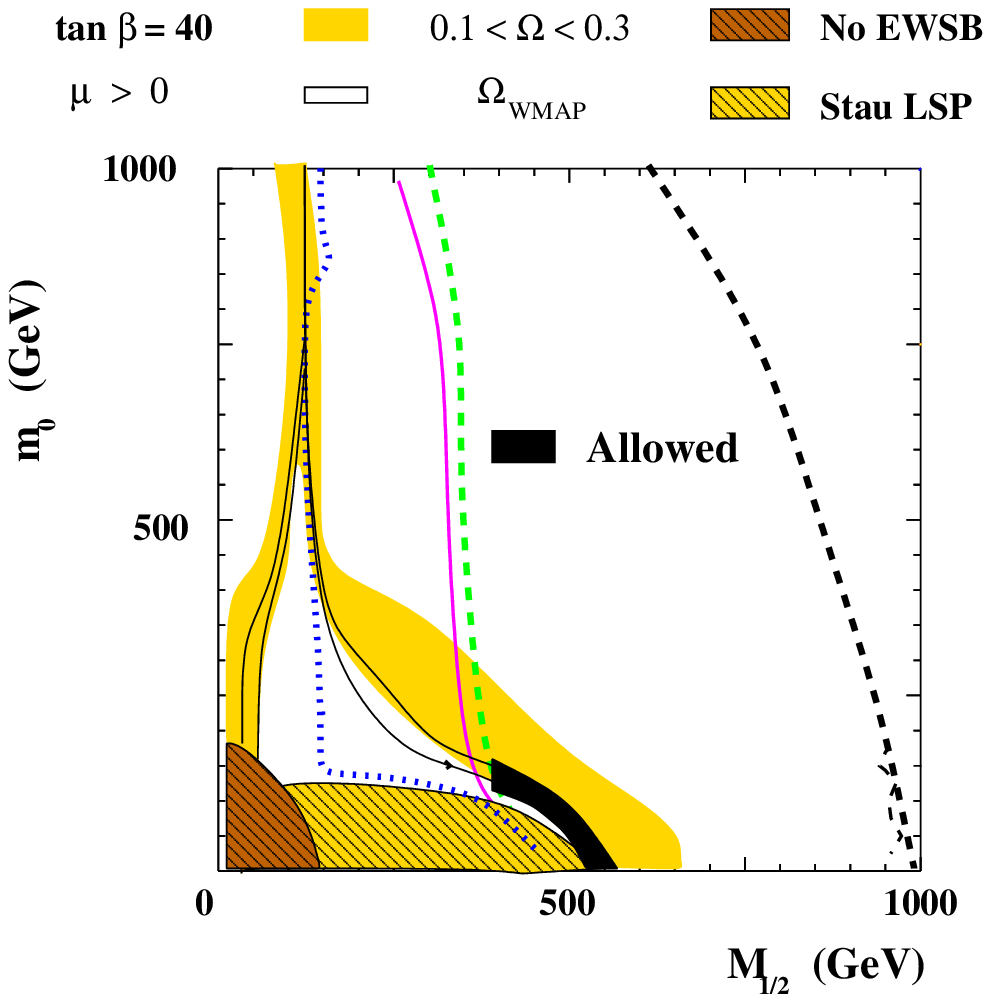}
\end{center}
\caption{Coannihilation region for $\tan\beta$=40, $A_0=0$ for 
CMSSM and $SU(5)_{RN}$.} 
\label{tb40}
\end{figure}

Looking at the peculiar $\surn$ phenomenology
of coannihilation, it is possible to understand the reason why there are no regions of the 
$\surn$ parameter space which give a neutralino
relic density compatible with WMAP for low values of $\tan\beta$. 
In Fig.~\ref{tb40} the plane ($m_0$, $M_{1/2}$) is plotted in the case of both CMSSM and $\surn$,
for $\tan\beta = 40$ and $A_0=0$.  
The WMAP allowed region is due to efficient $\stau$ coannihilation in both cases.
Two features are evident from such a comparison between the
CMSSM and the $SU(5)_{RN}$: in the case of $\surn$, 
(i) the WMAP compatible region
is much smaller and (ii) the shape of the allowed region is quite different. 
In fact, the allowed region cuts-off for a value of 
$M_{1/2}~\simeq~520$ GeV. This would correspond
to a LSP mass of around $240$ GeV; 
this corresponds to an \textit{upper bound} on 
the LSP mass. Therefore, in the
considered case of $\tan\beta$=40, $A_0=0$, the LSP mass can
achieve a very limited range of values 
($160~ {\rm GeV}\lesssim m_{\nt_1} \lesssim 240~ {\rm GeV}$)
which could be useful in distinguishing the model at colliders.  

These peculiar features of $\surn$ DM can be traced to the enhancement of the 
$\stau_R$ mass discussed above. A rough approximation for
the lightest $\stau$ mass is:
\begin{equation}
m^2_{\stau_1} \approx m_{\stau_R}^2 - m_\tau \mu \tan\beta
\label{mstau1}
\end{equation}
From here we can see that the term $ m_\tau \mu \tan\beta$, which
corresponds to the L-R mixing term of the $\stau$ mass matrix, is
crucial in setting the condition $m_{\stau_1}\simeq m_{\nt_1}$
for an efficient neutralino-stau coannihilation. As a consequence
of the GUT enhancement of $m_{\stau_R}^2$ in Eq.~(\ref{pregut}),
higher values of $\tan\beta$ are needed to lower the $\stau_1$ mass 
at the level of $ m_{\nt_1}$. This explains the lower limit
for $\tan\beta$ of Eq.~(\ref{lowerbounds}). Moreover, being the 
enhancement directly dependent on $M_{1/2}$, there is a value of
$M_{1/2}$ such that the L-R mixing term is no more able to lower 
sufficiently $m_{\stau_1}$ and the resultant coannihilation 
cross-section results too low to give the correct relic density\footnote{As discussed 
in the previous section, also $\mu \simeq -m^2_{H_u} - M^2_Z /2$ is enhanced in $\surn$, 
but not enough to contrast
the enhancement of $m_{\stau_R}^2$ in Eq.~(\ref{mstau1}), being the L-R term suppressed
by the small $\tau$ mass.}. 
From another point of view, the very peculiar shape of the
$\stau$ coannihilation region in $\surn$ is related to the fact
that the $\stau$-LSP region, along which the allowed region runs, 
is consistently reduced and bounded from above in $M_{1/2}$.
This is again due to the GUT enhancement of $m_{\stau_R}^2$. 

\begin{figure}[t]
\begin{center}
\includegraphics[width=0.36\textwidth]{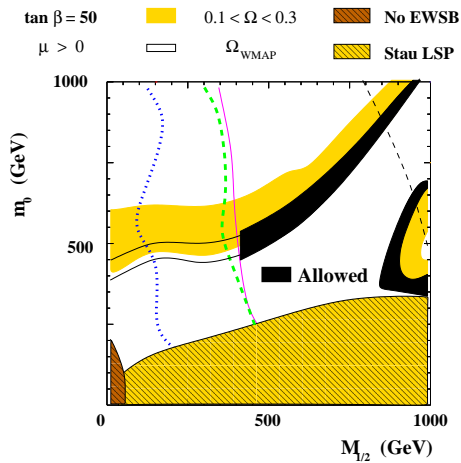}
\end{center}
\caption{WMAP allowed regions for $\tan\beta$=50, $A_0=0$ for $SU(5)_{RN}$.} 
\label{tb50}
\end{figure}

Let's now consider what happens to the other two DM branches
of CMSSM listed in the introduction.
The A-pole funnel region makes its appearance for large $\tan\beta ~\simeq~ 45-50$,
as in CMSSM.
In Fig.~\ref{tb50}, the funnel region is plotted for $\tan\beta$ = 50, 
$A_0=0$. 
We can also see large regions of the parameter space where the lightest 
stau is the LSP and thus regions of coannihilation also which 
are fused with the funnel region. As a consequence, the upper bound
on the LSP mass is no longer present for $\tb=50$, as it is also evident from 
Fig.~\ref{tb-mchi}.

As discussed above, the REWSB is very easy to be achieved due to the presence of additional
GUT effects and top-like Yukawa contribution from the neutrino Yukawa 
coupling to the up type Higgs between $M_X$ and $M_R$. 
The same effect induces an increasing of $\mu\simeq -m^2_{H_u} - M^2_Z /2$.
As a result, there is no focus-point region  
(at least up to $(m_0,\,M_{1/2}) \simeq 5\, {\rm TeV}$).
 
\begin{figure}[t]
\begin{center}
\includegraphics[width=0.36\textwidth]{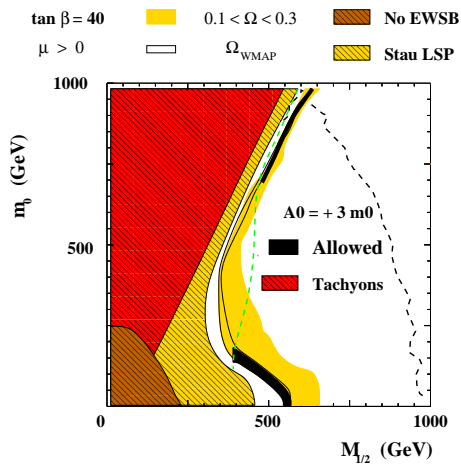}
\end{center}
\caption{$A_0$ = 3\, $m_0$ case: the $(m_0,M_{1/2})$ plane plot for $\tan\beta=40$; 
we can see the two branches of the stau coannihilation region.} 
\label{a0plus3}
\end{figure}

Finally, it is worth to mention that large values of $A_0$ can open an additional coannihilation
branch with respect to the one of Fig.~\ref{tb40}. The reason is that $m^2_{\stau_R}$ is suppressed,
even to negative values, by an additional A-term running effect. This effect is showed in 
Fig.~\ref{a0plus3}, for $A_0 = 3\,m_0$ and $\tb = 40$. Here a further thin line where coannihilation
provides the correct relic density is present and, as a consequence, there is no upper bound 
on $m_{\nt_1}$.

\section{Conclusions}
We have seen how GUT running effects and/or the presence of RH neutrinos can destabilize the
peculiar relations among parameters, which are needed in the CMSSM in order to provide a DM
relic density in accord with WMAP. In $\surn$, a major consequence is a severe constraint on 
the allowed 
range of $\tb$ ($\gtrsim 35$). Moreover, the peculiar phenomenology of $\stau$ coannihilation
region determines an upper bound on the LSP mass (around 250-350 GeV) for some regions of
the parameter space. Finally, the A-pole funnel branch appears for very large $\tb$, such as in
CMSSM, while focus point is absent.  

An interesting point to address is the possibility of distinguishing $\surn$ from 
CMSSM at colliders. The direct measurements of the LSP mass and $\tb$, maybe possible
at the LHC, can test the constrained ranges of such parameters allowed in $\surn$.  
Another interesting possibility is the study of the average polarization of $\tau$ leptons
coming from $\stau$ decays, which is possible with good accuracy at an International Linear Collider (ILC)
\cite{rohini}.
$\tau$-polarization gives a deep insight of the mixing structure of staus and neutralinos 
and should be able to distinguish $\surn$ from CMSSM as long as SUSY spectrum lies in the stau coannihilation
region \cite{ours2}.

\begin{center}
\textbf{Acknowledgments} 
\end{center}
\noindent I wish to thank the organizers of SUSY07 and
A. Faccia, R. Godbole, Y. Mambrini A. Masiero and S. K. Vempati for
collaborations on which this talk is based.
I would like to acknowledge support of the foundation ``Angelo Della Riccia'', 
and of the spanish MEC and FEDER under grant FPA2005-01678.

\end{document}